\documentclass{article}

\usepackage{amssymb}

\begin{document}

\title{\bf Noncommutativity, generalized uncertainty principle and FRW cosmology}
\author{A. Bina$^{1}$\thanks{email: a-bina@arshad.araku.ac.ir},\,\,\
K. Atazadeh $^2$\thanks{email: k-atazadeh@sbu.ac.ir}\,\ and S.
Jalalzadeh$^2$\thanks{email: s-jalalzadeh@sbu.ac.ir}
\\ $^1${\small Department of Physics, Arak
University, Arak, Iran}\\$^2${\small  Department of Physics, Shahid
Beheshti University, Evin, Tehran 19839, Iran}}
\date{\today}
\maketitle
\begin{abstract}
We consider the effects of noncommutativity and the generalized
uncertainty principle on the FRW cosmology with a scalar field. We
show that, the cosmological constant problem and removability of
initial curvature singularity find natural solutions in this
scenarios.
\vspace{5mm}\noindent
\end{abstract}
\section{Introduction}
Noncommutativity between spacetime coordinates which was first
introduced in \cite{1}, has been attracting considerable attention
in the recent past \cite{{2},{3},{4}}. This renewed interest has its roots
in the development of string and M-theories, \cite{{5},{6}}. However, in
all fairness, investigation of noncommutative theories may also be
justified in its own right because of the interesting predictions
regarding, for example, the IR/UV mixing and non-locality \cite{7},
Lorentz violation \cite{8} and new physics at very short distance
scales \cite{{9},{11}}. The impact of noncommutativity in cosmology has
also been considerable and  addressed in different works
\cite{12}. Hopefully, noncommutative cosmology would lead us to the
formulation of semiclassical approximations of quantum gravity and
tackles the cosmological constant problem \cite{13a}. Also it may
solve the compactification of extra dimensions \cite{{13a},{13b}}. To
study the effects of noncommutativity in cosmology one can use two
interesting different deformation of Poisson brackets, the Moyal
product and the Generalized Uncertainty Principle (GUP), see for
example \cite{{setare1},{setare2}}. The dynamical variable in general
theory of relativity is the spacetime itself. Unfortunately,
deformation of general relativity in noncommutative spacetime is
a difficult task to analyze even simple models. However, to have a
feeling of the effect of noncommutativity in cosmology, a proposal
was made by considering directly a noncommutativity of
minisuperspace of underling effective action. In this paper, as a
simple toy model, we will assume noncommutativity among
the gravitational and  scalar fields in classical relativity using both
of the above mentioned deformations. We will show that introducing
noncommutativity with the Moyal product may solve the cosmological
constant problem. On the other hand, GUP approach can remove the
initial Big-Bang curvature singularity.

\section{Commutative classical cosmology}
We consider a cosmological model in which the spacetime is assumed
to be of FRW type. Thus the corresponding metric can be written as
\begin{eqnarray}\label{2,1}
ds^2=-dt^2+\frac{R^2(t)}{\left(1+\frac{k}{4}r^2\right)^2} (dr^2+r^2
d\Omega^2)
    \hspace{.15cm},
\end{eqnarray}
where $k=1,0,-1$ denotes the usual spatial curvature and $R(t)$ is
the scale factor of the universe.

Let us start from the Einstein-Hilbert action plus a scalar field as
\begin{eqnarray}\label{2,2}
{\cal S}=\frac{1}{2\kappa^{2}}\int d^{4}x\sqrt{-g}{\cal R}+\int
d^{4}x\sqrt{-g}\left(-\frac{1}{2}(\nabla\phi)^{2}-U(\phi)\right)+{\cal
S}_{YGH},
\end{eqnarray}
where $\kappa=8\pi G$, ${\cal R}$ is Ricci scalar and ${\cal
S}_{YGH}$ is the York-Gibbons-Hawking boundary term. The second term
in equation (\ref{2,2}) shows the usual scalar field action
functional. Using the FRW line element (\ref{2,1}) the action
becomes
\begin{eqnarray}\label{2,3}
{\cal S}=\int
dt\left[-3R\dot{R}^{2}+3kR+(\frac{1}{2}\dot{\phi}^{2}-U(\phi))R^{3}\right].
\end{eqnarray}
In this paper, we will consider the case $k=0$ and
$U(\phi)=\Lambda$ such that the model describes a free scalar field with a
cosmological constant $\Lambda$. The underlying mechanical analogue
becomes more transparent if we make the following transformation
\cite{tucker}
\begin{eqnarray}\label{2,4}
\left\{
\begin{array}{lll}
x_{1}=R^{3/2}\cosh(\alpha\phi),\\
\\
x_{2}=R^{3/2}\sinh(\alpha\phi),
\end{array}
\right.
\end{eqnarray}
where $-\infty<\phi<\infty$, $0\leq R<\infty$ and
$\alpha^{2}=\frac{3}{8}$. Using the above transformations, the
minisuperspace Lagrangian is given by
\begin{eqnarray}\label{2,5}
{\cal
L}=\dot{x_{1}}^{2}-\dot{x_{2}}^{2}+2\alpha^{2}\Lambda(x_{1}^{2}-x_{2}^{2}).
\end{eqnarray}
Now, we can write the effective Hamiltonian as
\begin{eqnarray}\label{2,6}
{\cal H}=\frac{1}{4}(p_{1}^{2}-p_{2}^{2})-
\omega^2(x_{1}^{2}-x_{2}^{2}),
\end{eqnarray}
where $p_{i}$ denotes the conjugate momenta corresponding to $x_{i}$ and
$\omega^{2}=2\alpha^{2}\Lambda$. Using the following commutative relation
between $x_{i}$ and $p_{i}$ 
\begin{eqnarray}\label{2,7}
\{x_{i},p_{j}\}=\delta_{ij},~~~~~~~i, j=1, 2.
\end{eqnarray}
the canonical equations of motion become
\begin{eqnarray}\label{2,8}
\left\{
\begin{array}{lll}
\dot{x}_{1}=\frac{1}{2}p_{1}, ~~~~~\dot{x}_{2}=-\frac{1}{2}p_{2},
\\
\dot{p_{1}}=2\omega^{2}x_{1}, ~~~~~\dot{p_{2}}=-2\omega^{2}x_{2},
\end{array}
\right.
\end{eqnarray}
where a dot denotes derivative with respect to $t$. Then we have the differential 
equations of motion
\begin{eqnarray}\label{2,9}
\ddot{x_{i}}-\omega^{2}x_{i}=0,
\end{eqnarray}
with solutions
\begin{eqnarray}\label{2,10}
x_{i}(t)=A_{i}e^{\omega t}+B_{i} e^{-\omega t},
\end{eqnarray}
where $A_{i}$ and $B_{i}$ are integration constants. Note that the
Hamiltonian constraint (${\cal H}=0$) imposes the following relation
on these constants as
\begin{eqnarray}\label{2,11}
A_{1} B_{1}- A_{2} B_{2}=0.
\end{eqnarray}
Finally, using  equation (\ref{2,11}) and choosing
$A_{1}=B_{2}$ and $A_{2}=B_{1}$ , the scale factor and scalar field
are given by
\begin{eqnarray}\label{2,12}
\left\{
\begin{array}{lll}
R(t)^{3}=2(A^{2}_{1}-A^{2}_{2})\sinh2\omega t,\\
\\
\phi(t)=\frac{1}{\alpha}\tanh^{-1}\left(\frac{A_{2}e^{\omega
t}+A_{1}e^{-\omega t}}{A_{1}e^{\omega t}+A_{2}e^{-\omega t}}\right).
\end{array}
\right.
\end{eqnarray}
Note that if $\omega^{2} = \frac{3}{4} \Lambda<0$, the hyperbolic
functions are replaced by their trigonometric counterparts in the
above solutions.

\section{Moyal product approach}
We now concentrate on the noncommutativity concepts with Moyal
product in phase space. The Moyal product in phase space may be
traced to an early intuition by Wigner \cite{wig} which has been
developing over the past decades \cite{gozzi,smail,bertolami}.
Noncommutativity in classical physics \cite{15} is described by the
Moyal product law between two arbitrary functions of positions and
momenta as
\begin{eqnarray}\label{3,1}
(f\star_\alpha
g)(x)=\exp\left[\frac{1}{2}\alpha^{ab}\partial_a^{(1)}\partial_b^{(2)}\right]
f(x_1)g(x_2)|_{x_1=x_2=x},
\end{eqnarray}
such that
\begin{eqnarray}\label{3,2}
\alpha_{ab}=\left(%
\begin{array}{cc}
\theta_{ij}  & \delta_{ij}+\sigma_{ij} \\
-\delta_{ij}-\sigma_{ij} & \bar{\theta}_{ij}\\
\end{array}%
\right),
\end{eqnarray}
where the $N\times N$ matrices $\theta$ and $\bar{\theta}$ are
assumed to be antisymmetric with $2N$ being the dimension of the
classical phase space and $\sigma$ can be written as a combination
of $\theta$ and $\bar{\theta}$. With this product law, the deformed
Poisson brackets can be written as
\begin{eqnarray}\label{3,3}
\{f,g\}_\alpha=f\star_\alpha g-g\star_\alpha f.
\end{eqnarray}
A simple calculation shows that
\begin{eqnarray}\label{3,4}
\begin{array}{lll}
\{x_{i},x_{j}\}_\alpha=\theta_{ij}, \\
\{x_{i},p_{j}\}_\alpha=\delta_{ij}+\sigma_{ij}, \\
\{p_{i},p_{j}\}_\alpha=\bar{\theta}_{ij} \hspace{.15cm}.
\end{array}
\end{eqnarray}
It is worth noting at this stage that, in addition to the
noncommutativity in $(x_1,x_2)$,  we have also considered
noncommutativity in the corresponding momenta. This should be
interesting since its existence  is due to essentially to the
existence of noncommutativity in the space sector \cite{gozzi,15}
and it would somehow be natural to include it in our considerations.

To proceed, we consider the following transformations on the
classical phase space $(x_{i},p_{j})$
\begin{eqnarray}\label{3,5}
\begin{array}{ll}
{x_{i}'}=x_{i}-\frac{1}{2}\theta_{ij} p^{j}, \\
{p_{i}'}=p_{i}+\frac{1}{2}\bar{\theta}_{ij} x^{j}
 .
\end{array}
\end{eqnarray}
It is easy to check  that if the $(x_{i},p_{j})$ obey the usual
Poisson algebra (\ref{2,7}), then
\begin{eqnarray}\label{3,6}
\begin{array}{lll}
\{{x_{i}'},{x_{j}'}\}_{P}=\theta_{ij},\\
\{{x_{i}'},{p_{j}'}\}_{P}=\delta_{ij}+\sigma_{ij},\\
\{{p_{i}'},{p_{j}'}\}_{P}=\bar{\theta}_{ij}.
\end{array}
\end{eqnarray}
These commutation relations are the same as (\ref{3,4}).
Consequently, to introduce noncommutativity, it is more
convenient to work with Poisson brackets (\ref{3,6}) than
$\alpha$-star deformed Poisson brackets (\ref{3,4}). It is important
to note that the relations represented by equations (\ref{3,4}) are
defined in the spirit of the Moyal product given above. However, in
the relations defined in (\ref{3,6}), the variables $(x_{i},p_{j})$
obey the usual Poisson bracket relations so that the two sets of
deformed and ordinary Poisson brackets represented by relations
(\ref{3,4}) and (\ref{3,6}) should be considered as distinct.

Let us change the commutative Hamiltonian (\ref{2,6}) with minimal
variation to
\begin{eqnarray}\label{3,7}
{\cal
H'}=\frac{1}{4}(p'^{2}_{1}-p'^{2}_{2})-\omega^2(x'^{2}_{1}-x'^{2}_{2}),
\end{eqnarray}
where we have the commutation relations
\begin{eqnarray}\label{3,8}
\begin{array}{lll}
\{{x_{i}'},{x_{j}'}\}_{P}=\theta \epsilon_{ij},\\
\{{x_{i}'},{p'_{j}}\}_{P}=(1+\sigma)\delta_{ij},\\
\{{p_{i}'},p_{j}'\}_{P}=\bar{\theta} \epsilon_{ij},
\end{array}
\end{eqnarray}
with $\epsilon_{ij}$ being a totally anti-symmetric tensor and
$\sigma$ is given by
\begin{eqnarray}\label{3,9}
\sigma=\frac{1}{4}\bar{\theta}\theta.
\end{eqnarray}
We have also set $\theta_{ij}=\theta \epsilon_{ij}$ and
$\bar{\theta}_{ij}=\bar{\theta} \epsilon_{ij}$. Using the
transformation of equation (\ref{3,5}), Hamiltonian (\ref{3,7})
becomes
\begin{eqnarray}\label{3,10}
{\cal
H}=\frac{1}{4}\left(1+\omega^2\theta^2\right)(p_{1}^{2}-p_{2}^{2})-\left(\omega^2+\frac{\bar{\theta}^{2}}{16}\right)
(x_{1}^{2}-x_{2}^{2})
+\left(\frac{\bar{\theta}}{4}+\theta\omega^2\right)(p_{1}x_{2}+p_{2}x_{1}).
\end{eqnarray}
The equations of motion corresponding to the Hamiltonian
(\ref{3,10}) are given by
\begin{eqnarray}\label{3,11}
\left\{
\begin{array}{lll}
\dot{x_{1}}=\{x_{1},{\cal
H}\}_{P}=\frac{1}{2}\left(1+\omega^2\theta^2\right)p_{1}+
\left(\frac{\bar{\theta}}{4}+\theta\omega^2\right)x_{2},\\
\\
\dot{x_{2}}=\{x_{2},{\cal
H}\}_{P}=\frac{-1}{2}\left(1+\omega^2\theta^2\right)p_{2}+
\left(\frac{\bar{\theta}}{4}+\theta\omega^2\right)x_{1},\\
\\
\dot{p_{1}}=\{p_{1},{\cal
H}\}_{P}=2\left(\omega^2+\frac{\bar{\theta}^{2}}{16}\right)x_{1}-
\left(\frac{\bar{\theta}}{4}+\theta\omega^2\right)p_{2},\\
\\
\dot{p_{2}}=\{p_{2},{\cal
H}\}_{P}=-2\left(\omega^2+\frac{\bar{\theta}^{2}}{16}\right)x_{2}-
\left(\frac{\bar{\theta}}{4}+\theta\omega^2\right)p_{1},
\end{array} \right.
\end{eqnarray}
where we have used relations (\ref{2,7}). Again It can  be easily
checked that if one writes the equations of motion for
noncommutative variables, equations (\ref{3,8}), with respect to the
Hamiltonian (\ref{3,7}) and  uses transformation rules (\ref{3,5}),
one gets a linear combination of the equations of motion
(\ref{3,11}). This points to the fact that these two approaches are
equivalent. Now, as a consequence of equations of motion
(\ref{3,11}), one has
\begin{eqnarray}\label{3,12}
\left\{
\begin{array}{lll}
\ddot{x}_{1}-2\left(\frac{\bar{\theta}}{4}+\theta\omega^2\right)\dot{x}_{2}
-\left(1-\frac{\theta\bar{\theta}}{4}\right)^{2}\omega^{2}x_{1}=0,
\\
\ddot{x}_{2}-2\left(\frac{\bar{\theta}}{4}+\theta\omega^2\right)\dot{x}_{1}
-\left(1-\frac{\theta\bar{\theta}}{4}\right)^{2}\omega^{2}x_{2}=0.
\end{array}
\right.
\end{eqnarray}
Note that upon setting $\theta=\bar{\theta}=0$, we get back equation
(\ref{2,9}). Solutions of the equations (\ref{3,12}) can be written
as follows
\begin{eqnarray}\label{3,14}
\left\{
\begin{array}{lll}
x_1(t)=K_{1}e^{(\eta+\xi)t}+K_{2}e^{(\eta-\xi)t}+
K_{3}e^{-(\eta-\xi)t}+K_{4}e^{-(\eta+\xi)t}, \\
\\
x_2(t)=K_{1}e^{(\eta+\xi)t}+K_{2}e^{(\eta-\xi)t}-
K_{3}e^{-(\eta-\xi)t}-K_{4}e^{-(\eta+\xi)t}, \\
\end{array}
\right.
\end{eqnarray}
where
\begin{eqnarray}\label{3,15}
\eta=\theta\omega^2+\frac{\bar{\theta}}{4}\hspace{3mm}\mbox{,}\hspace{3mm}
\xi=(\eta^{2}+\lambda)^{1/2}\hspace{3mm}\mbox{and}\hspace{3mm}
\lambda=\omega^{2}\left(1-\frac{\theta\bar{\theta}}{4}\right)^{2},
\end{eqnarray}
with $K_{1}$, $K_{2}$, $K_{3}$  and $K_{4}$ being the constants of
integration. The Hamiltonian constraint, ${\cal H}'=0$, leads to
\begin{eqnarray}\label{3,16}
K_{3}+K_{4}=0~~~~~~or~~~~~ K_{1}=K_{2}=0 ,
\end{eqnarray}
Thus, from equations (\ref{2,4}) and (\ref{3,14}) we can recover the
scale factor and scalar field $\phi(t)$ as
\begin{eqnarray}\label{3,17}
\left\{
\begin{array}{lll}
R(t)^{3}= e^{-2\xi t} \left[K_{1}e^{2\xi
t}+K_{2}\right]\left[K_{3}e^{2\xi
t}+K_{4}\right], \\
\\
\
\phi(t)=\frac{1}{\alpha}\tanh^{-1}\left(\frac{K_{1}e^{(\eta+\xi)t}+K_{2}e^{(\eta-\xi)t}-
K_{3}e^{-(\eta-\xi)t}-K_{4}e^{-(\eta+\xi)t}}{K_{1}e^{(\eta+\xi)t}+K_{2}e^{(\eta-\xi)t}+
K_{3}e^{-(\eta-\xi)t}+K_{4}e^{-(\eta+\xi)t}}\right).
\end{array}
\right.
\end{eqnarray}
It is clear from the solutions (\ref{3,17}),  that we still
have a late time de Sitter phase solution
\begin{eqnarray}\label{3,17a}
R^3 = K_1K_3 e^{2\xi t},
\end{eqnarray}
and an initial Big-Bang singularity. However, equation (\ref{3,17a})
dictates the new cosmological constant in the noncommutative case
as\begin{eqnarray}\label{3,17b}
\Lambda_{nc} =
\frac{4}{3}\left(1+\theta^2\omega^2\right)\left(\omega^2+\frac{1}{16}\bar{\theta}^2\right)
=
(1+\frac{3}{4}\theta^2\Lambda)(\Lambda+\frac{1}{12}\bar{\theta}^2).
\end{eqnarray}
The effective cosmological constant on the other hand, is a measure
of the ultraviolet cutoff in the theory. Hence, the above equation
can be interpreted as a redefinition of the cutoff in the theory due
the presence of noncommutativity. In this regard, according to
\cite{darabi} we take
\begin{eqnarray}
\Lambda \sim M_{EW}^4, \hspace{1cm} \theta^2 \sim M_{EW}^{-4},
\hspace{1cm}and \hspace{1cm} \bar{\theta}^2 \sim M_{P}^4,
\end{eqnarray}
where $M_{EW}$ is the electroweak mass scale. Therefore, from the
above relations we obtain
\begin{eqnarray}
\Lambda_{nc}=M^4_{P},
\end{eqnarray}
which  defines the cutoff in the noncommutative model. In  other
words if we assume $M_{EW}$ to be the natural cutoff in the
original commutative model, the Planck mass is a cutoff in the
noncommutative model. A similar discussion can be found for a 
classical cosmological model in \cite{13a} and at the quantum level
in \cite{darabi}.

\section{Generalized Uncertainty Principle (GUP) approach}
According to \cite{chang1}, the modified commutation relations in
GUP approach for more than one dimensional systems is given by
\begin{eqnarray}\label{4,1}
\{x'_{i},p'_{j}\}=\delta_{ij}(1+\beta p'^{2})+\beta'p'_{i}p'_{j},
\end{eqnarray}
where $\beta$ and $\beta'$ are constant. If we assume the components
of momenta, $p'_{i}$, to commute with each other i.e.
\begin{eqnarray}\label{4,2}
\{p'_{i},p'_{j}\}=0,
\end{eqnarray}
then the commutation relations among the coordinates $x'_{i}$ are
almost uniquely determined by the Jacobi identity as
\cite{kempf1,kempf2}
\begin{eqnarray}\label{4,3}
\{x'_{i},x'_{j}\}=\frac{(2\beta-\beta')+(2\beta+\beta')\beta
p'^{2}}{1+\beta p'^{2}}(p'_{i}x'_{j}-p'_{j}x'_{i}).
\end{eqnarray}
Similar to the previous section, we can introduce the following
transformations
\begin{eqnarray}\label{4,4}
\left\{
\begin{array}{lll}
x'_{i}=\left[(1+\beta p^{2})x_{i}+
\beta' p_{i}p_{j}x_{i}+\gamma p_{i}\right], \\
\\
p'_{i}=p_{i}.
\end{array}
\right.
\end{eqnarray}
One can show that if $(x_{i},p_{j})$ obey the ordinary Poisson
algebra (\ref{2,7}), then $(x'_{i},p'_{j})$ satisfy the deformed
commutation relation (\ref{4,1}) with the usual Poisson brackets.
The arbitrary constant $\gamma$ in the representation of $x'_{i}$
does not affect the commutation relation among the $x'_{i}$'s and is given by
\cite{chang2}
\begin{eqnarray}\label{4,5}
\gamma=\beta+\beta'\left(\frac{D+1}{2}\right).
\end{eqnarray}
Note that for $D=2$, we have $\gamma=\beta+\frac{3}{2}\beta'$ and
$p^{2}=p^{2}_{1}+p^{2}_{1}$. Moreover, $\beta$ and
$\beta'>0$ and are
assumed small up to the first order.

The minimal variation of  Hamiltonian (\ref{2,6}) in the GUP approach
then becomes
\begin{eqnarray}\label{4,6}
{\cal
H'}=\frac{1}{4}(p'^{2}_{1}-p'^{2}_{2})-\omega^2(x'^{2}_{1}-x'^{2}_{2})=
\frac{1}{4}(p^{2}_{1}-p^{2}_{2})-\\\nonumber \omega^2\{(1+2\beta
p^{2})(x^{2}_{1}-x^{2}_{2})+2(x_{1}p_{1}-x_{2}p_{2})[\gamma+\beta'(x_{1}p_{1}+x_{2}p_{2})]\}.
\end{eqnarray}
The equations of motion corresponding to the above modified
Hamiltonian (\ref{4,6}) are given by
\begin{eqnarray}\label{4,7}
\left\{
\begin{array}{lll}
\dot{x}_{1}=\frac{p_{1}}{2}-\omega^{2}[4\beta
(x^{2}_{1}-x^{2}_{2})p_{1}+2\beta'(x_{1}p_{1}-x_{2}p_{2})x_{1}+2(\gamma+\beta'(x_{1}p_{1}+x_{2}p_{2}))x_{1}], \\
\\
\ \dot{x}_{2}=\frac{-p_{2}}{2}-\omega^{2}[4\beta
(x^{2}_{1}-x^{2}_{2})p_{2}+2\beta'(x_{1}p_{1}-x_{2}p_{2})x_{2}-2(\gamma+\beta'(x_{1}p_{1}+x_{2}p_{2}))x_{2}],\\
\\
\dot{p}_{1}=\omega^{2}\{2(1+2\beta
p^{2})x_{1}+2\beta'(x_{1}p_{1}-x_{2}p_{2})p_{1}+2(\gamma+\beta'(x_{1}p_{1}+x_{2}p_{2}))p_{1}],\\
\\
\dot{p}_{2}=\omega^{2}\{-2(1+2\beta
p^{2})x_{2}+2\beta'(x_{1}p_{1}-x_{2}p_{2})p_{2}-2(\gamma+\beta'(x_{1}p_{1}+x_{2}p_{2}))p_{2}].
\end{array}
\right.
\end{eqnarray}
After a little calculation, we obtain
\begin{eqnarray}\label{4,8}
\left\{
\begin{array}{lll}
\ddot{x}_{1}=\omega^{2}x_{1}-8\omega^{4}(\beta+\beta')x_{1}^{3}+8\omega^{4}\beta
x_{1}x_{2}^{2}-8\omega^{2}(\beta+\beta')x_{1}\dot{x}_{1}^{2}+16\omega^{2}\beta
x_{2}\dot{x}_{1}\dot{x}_{2}+8\omega^{2}\beta x_{1}\dot{x}_{2}^{2},
\\
\\
\ddot{x}_{2}=\omega^{2}x_{2}-8\omega^{4}\beta'x_{2}^{3}-8\omega^{2}\beta(x_{1}^{2}-x_{2}^{2})\dot{x}_{1}+8\omega^{2}\beta
x_{2}\dot{x}_{1}^{2}+16\omega^{2}\beta
x_{1}\dot{x}_{1}\dot{x}_{2}-8\omega^{2}(\beta+\beta')x_{2}\dot{x}_{2}^{2}.
\end{array}
\right.
\end{eqnarray}
At this stage, we  solve  equations (\ref{4,8}) using
the perturbation method to expand new solutions around commutative
solutions as
\begin{eqnarray}\label{4,10}
x_i = x_{0i}+\beta f_i(t)+\beta ' g_i(t),
\end{eqnarray}
where $x_{0i}$ obey the commutative solutions.  After substituting the expansion (\ref{4,10})
into  equations  (\ref{4,8}) and ignoring higher orders of
$\beta$, $\beta'$ and $\omega^2$ terms we obtain
\begin{eqnarray} \left\{
\begin{array}{lll}
\ddot{f}_i &=& \omega^2 f_i, \\
\ \ddot{g}_i &=& \omega^2 g_i.
\end{array}
\right.
\end{eqnarray}\label{4,11}
Now inserting the solutions of above equations into (\ref{4,10}) and
using the commutative solutions (\ref{2,10}), we obtain
\begin{eqnarray} \left\{
\begin{array}{lll}
x_1 = (A_1 + \beta M_1 + \beta' M_3)e^{\omega t} + (A_2 + \beta M_2 + \beta' M_4)e^{-\omega t}, \\
\\
x_2 = (A_2 + \beta M_5 + \beta' M_7)e^{\omega t} + (A_1 + \beta M_6
+ \beta' M_8)e^{-\omega t}.
\end{array}
\right.
\end{eqnarray}\label{4,12}
On the other hand the Hamiltonian constraint (\ref{4,6}) gives the
following relations between the integration constants
\begin{eqnarray} \left\{
\begin{array}{lll}
A_1 (M_2 - M_5) + A_2( M_1 - M_6)=0 , \\
\\
A_1 (M_4 - M_7) + A_2( M_3 - M_8)=0 .
\end{array}
\right.
\end{eqnarray}\label{4,13}
Finally, using equation (\ref{2,4}) the scale factor and scalar
field become
\begin{eqnarray}\label{4,14}
R(t)^3 &=& (A_1^2 - A_2^2) \sinh(2\omega t) + 2[
\beta(A_1M_1-A_2M_5) + \beta'(A_1 M_3 - A_2M_7) ]e^{2\omega
t}\\\nonumber &+& 2[ \beta(A_2M_2-A_1M_6) + \beta'(A_2 M_4 - A_1M_8)
]e^{-2\omega t},
\end{eqnarray}
and
\begin{eqnarray}\label{4,15}
\phi(t)=\frac{1}{\alpha}\tanh^{-1}\left[\frac{(A_2 + \beta M_5
+\beta' M_7)e^{\omega t} + (A_1 + \beta M_6 + \beta' M_8)e^{-\omega
t}}{(A_1 + \beta M_1 + \beta' M_3)e^{\omega t} + (A_2 + \beta M_2 +
\beta' M_4)e^{-\omega t}}\right].
\end{eqnarray}
Solution (\ref{4,14}) shows that we have a late de Sitter phase,
like commutative and Moyal noncommutative cases, with unchanged
cosmological constant. Hence in the  GUP modification of this simple cosmological model, the cosmological constant remains unchanged. However, Solution (\ref{4,14}), dose
not have a curvature singularity  and it seems the GUP approach addresses
the initial Big-bang singularity problem.

\section{Conclusions}
In this manuscript we have introduced two different  deformation
methods between scale factor of the FRW universe and the scalar field in
minisuperspace. We have shown that the classical solutions of such
models clearly point to a possible resolution of the cosmological
constant problem  using the  Moyal product deformation and the initial Big-Bang
singularity wither GUP approach . It appears that noncommutative models in
minisuperspace in conjunction with of the both Moyal and GUP methods
\cite{Nozari} can clarify simultaneously the two above mentioned
problems.
\vspace{10mm}\noindent\\

\end{document}